\documentclass{llncs}

\usepackage{ifpdf}
\usepackage{color}               
\ifpdf
\usepackage{ae,aecompl}
\usepackage{graphicx}            
\else
\usepackage[dvips]{graphicx}
\DeclareGraphicsExtensions{.ps,.jpg,.jpeg,.eps,.pdf,.png,.mps}
\fi
\usepackage{picins}
\makeatletter
\def\piccaption{\@dblarg{\@piccaption}} 
\makeatother

\usepackage{amsmath,amssymb}
\usepackage{algorithm}
\usepackage{algorithmic}
\usepackage[latin1]{inputenc}
\usepackage[OT1]{fontenc}
\usepackage{comment}
\excludecomment{toolong}

\usepackage[all]{xy}

\newtheorem{theo}{Theorem}

\newenvironment{myproof}[1]{
\vspace{-5ex}

\hspace*{1em}\begin{trivlist}\item[\hskip\labelsep{\textsc{#1}}]}%
{\end{trivlist}}
\renewenvironment{proof}{
\begin{myproof}{Proof.}}{\hfill $\Box$\end{myproof}}

\newenvironment{proofclaim}{%
\vspace{-5ex}
\addtolength{\linewidth}{-1ex}
\begin{center}
\hspace*{1ex}\begin{trivlist}
\item\begin{myproof}{Proof of the claim.}}{
$\Diamond$\end{myproof}
\end{trivlist}
\end{center}
\addtolength{\linewidth}{1ex}

\vspace{-2ex}
}

\newcommand{\call}[1]{\ensuremath{{\mathcal{#1}}}}

\newcommand{\mathconstante}[1]{\ensuremath{{\mathrm{#1}}}}

\newcommand{\Variables}{\call{X}}

\newcommand{\Constants}{\mathconstante{C}}
\newcommand{\fonction}[2]{\ensuremath{{\mathconstante{#1}(#2)}}}

\newcommand{\set}[1]{\ensuremath{{\left\lbrace #1 \right\rbrace}}}

\newcommand{\Const}[1]{\fonction{Const}{#1}}
\newcommand{\termset}[1]{{\fonction{T}{#1}}} 
\newcommand{\gsig}[1]{\termset{#1}}
\newcommand{\vsig}[1]{\termset{#1,\Variables}}

\newcommand{\TGX}{\vsig{\call{G}}}

\newcommand{\rhnorm}[2]{\ensuremath{{(#1)\!\!\downarrow_{#2}}}}

\newcommand{\norm}[1]{\rhnorm{#1}{}}

\newcommand{\syncsub}[1]{\fonction{\text{Sub}_{\text{syn}}}{#1}}
\newcommand{\minic}{\mathconstante{c_{min}}}
\newcommand{\speC}{\mathconstante{C_{spe}}}

\newcommand{\Var}[1]{\fonction{Var}{#1}}
\newcommand{\Sub}[1]{\fonction{Sub}{#1}}

\newcommand{\Cons}[1]{\fonction{Cons}{#1}}
\newcommand{\atoms}[1]{\fonction{Atoms}{#1}}
\newcommand{\Atoms}{\mathconstante{Atoms}}

\newcommand{\factor}[1]{\fonction{Factors}{#1}}

\newcommand{\sig}[1]{\fonction{Sign}{#1}}
\newcommand{\head}[1]{\fonction{Sign}{#1}}

\newcommand{\Supp}[1]{\fonction{Supp}{#1}}

\newcommand{\tq}{\ensuremath{\,|\,}}
\newcommand{\condset}[2]{\set{#1\tq{}#2}}

\newcommand{\grule}{\mathconstante{L}}
\newcommand{\lrule}[1]{{\ensuremath{\grule^{#1}}}}
\newcommand{\glrule}[1]{{\ensuremath{\grule^{#1,\mathconstante{g}}}}}
\newcommand{\rhclos}[2]{{\ensuremath{\overline{#1}^{#2}}}}
\newcommand{\clos}[1]{\rhclos{#1}{}}

\newenvironment{decisionproblem}[1]{
\vspace*{-1ex}
\begin{tabbing}
  \underline{\textbf{#1}}\\
  \hspace*{2em}\= \textbf{Input:}~ \=}{
\end{tabbing}
\vspace*{-1em}
}

\newcommand{\change}{}
\newcommand{\entreeu}[1]{ \hbox{\vbox{\parbox[t]{0.8\linewidth}{#1}}}\\}
  
\newcommand{\sortie}[1]{
\> \textbf{Output:} \> \hbox{\vbox{\parbox[t]{0.8\linewidth}{#1}}}\\}

\newcommand{\unif}{\ensuremath{\stackrel{?}{=}}}
\newcommand{\intrus}[3]{\ensuremath{\left\langle #1,#2,#3\right\rangle}}
\newcommand{\nintrus}[1]{\intrus{\call{F}_{#1}}{S_{#1}}{\call{E}_{#1}}}

\newcommand{\IAU}{\ensuremath{\call{I}_{\mbox{\tiny\rm AU}}}}
\newcommand{\FAU}{\ensuremath{\call{F}_{\mbox{\tiny\rm AU}}}}
\newcommand{\SAU}{\ensuremath{S_{\mbox{\tiny\rm AU}}}}
\newcommand{\EAU}{\call{E}_{\mbox{\tiny\rm AU}}}
\newcommand{\Ifree}{\ensuremath{\call{I}_{\mbox{\tiny\rm free}}}}
\newcommand{\IH}{\ensuremath{\call{I}_{\mbox{\tiny\rm h}}}}
\newcommand{\IG}{\ensuremath{\call{I}_{\mbox{\tiny\rm g}}}}
\newcommand{\IF}{\ensuremath{\call{I}_{\mbox{\tiny\rm f}}}}
\newcommand{\FH}{\ensuremath{\call{F}_{\mbox{\tiny\rm h}}}}
\newcommand{\SH}{\ensuremath{S_{\mbox{\tiny\rm h}}}}
\newcommand{\EH}{\ensuremath{\call{E}_{\mbox{\tiny\rm h}}}}

\newcommand{\boite}[1]{
\begin{center}
  \fbox{\parbox{0.94\linewidth}{#1}}
\end{center}}

\newcounter{lemc}
\newcommand{\hop}{\ensuremath{\mathop{\mathrm{h}}}}
\newcommand{\hfunc}[1]{\ensuremath{\hop(#1)}}
\newcommand{\gop}{\ensuremath{\mathop{\mathrm{g}}}}
\newcommand{\gfunc}[2]{\ensuremath{\gop(#1_1,#1_2,#2_1,#2_2)}}
\newcommand{\fop}{\ensuremath{\mathop{\mathrm{f}}}}
\newcommand{\ffunc}[2]{\ensuremath{\fop(#1_1,#1_2,#2_1,#2_2)}}
\newcommand{\letters}[1]{\ensuremath{\mbox{\rm letters}(#1)}}
\newcommand{\Kappa}{\ensuremath{\mbox{\rm K}}}
\newcommand{\arity}[1]{\ensuremath{\mathop{\mbox{\rm ar}}(#1)}}
\newcommand{\topsym}[1]{\ensuremath{\mathop{\mbox{\rm top}}(#1)}}
\newcommand{\mode}[2]{\ensuremath{\mathop{\mbox{\rm m}}(#1,#2)}}
\newcommand{\card}[1]{\ensuremath{\mathop{\mbox{\rm card}}}}
\newcommand{\deduce}[2]{\ensuremath{#1\to #2}}

\begin{document}

\title{A Symbolic Intruder Model for Hash-Collision Attacks
\thanks{supported by ARA-SSIA Cops and ACI JC 9005}} \author{Yannick
Chevalier \and Mounira Kourjieh}
\institute{IRIT Universit{\'e} Paul Sabatier, France  \\
  email: $\lbrace$ychevali,kourjieh$\rbrace$@irit.fr }

\date{\today}

\maketitle 

\begin{abstract}
  In the recent years, several practical methods have been published to
  compute collisions on some commonly used hash functions. Starting from
  two messages $m_1$ and $m_2$ these methods permit to compute $m_1'$
  and $m_2'$ \textit{similar} to the former such that they have the
  same image for a given hash function. In this paper we present a
  method to take into account, at the symbolic level, that an intruder
  actively attacking a protocol execution may use these collision
  algorithms in reasonable time during the attack. This decision
  procedure relies on the reduction of constraint solving for an
  intruder exploiting the collision properties of hash functions to
  constraint solving for an intruder operating on words, that is with
  an associative symbol of concatenation. The decidability of the
  latter is interesting in its own right as it is the first
  decidability result that we are aware of for an intruder system for
  which unification is infinitary, and permits to consider in other
  contexts an associative concatenation of messages instead of their
  pairing.
\end{abstract}

\bibliographystyle{plain}

\sloppy 
\section{Introduction}

\paragraph{Hash functions.}  Cryptographic hash functions play a
fundamental role in modern cryptography. While related to
conventional hash functions commonly used in non-cryptographic
computer applications - in both cases, larger domains are mapped to
smaller ranges - they have some additional properties. Our focus is
restricted to cryptographic hash functions (hereafter, simply hash
functions), and in particular to their use as cryptographic primitive
for data integrity, authentication, key agreement, e-cash and many
other cryptographic schemes and protocols.  Hash functions take a
message as input and produce an output referred to either as a
\emph{hash-code}, \emph{hash-result}, or \emph{hash-value}, or simply
\emph{hash}.

\paragraph{Collisions.}  A hash function is many-to-one, implying that
the existence of collisions (pairs of inputs with the identical
output) is unavoidable.  However, only a few years ago, it was
intractable to compute collisions on hash functions, so they were
considered to be collision-free by cryptographers, and protocols were
built upon this assumption. From the nineties on, several authors have
proved the tractability of finding pseudo-collision and collision
attacks over several hash functions. Taking this into account, we
consider that cryptographic hash functions have the following
properties:
\begin{itemize}
\item the input can be of any length, the output has a fixed length,
  $\hfunc x$ is relatively easy to compute for any given $x$;
\item pre-image resistance: for essentially all pre-specified outputs,
  it is computationally infeasible to find any input which hashes to
  that outputs, i.e., to find any $x$ such that $y=\hfunc x$ when given
  $y$;
\item 2nd-pre-image resistance: it is computationally infeasible to
  find any second input which has the same output as any specified
  input, i.e., given $x$ , to find $x^{\prime}$ different from $x$
  such that $\hfunc x=\hfunc{x^{\prime}}$;
\item hash collision: it is computationally \emph{feasible} to compute two
  distinct inputs $x$ and $x^{\prime}$ which hash to the same output,
  i.e, $\hfunc x=\hfunc{x^{\prime}}$ provided that $x$ and
  $x^{\prime}$ are created at the same time and independently one of
  the other.
\end{itemize}
In other words, a collision-vulnerable hash function $h$ is one for
which an intruder can find two different messages $x$ and $x'$ with
the same hash value. To mount a collision attack, an adversary
would typically begin by constructing two messages with the same hash
where one message appears legitimate or innocuous while the other
serves the intruder's purposes.  For example, consider the following
simple protocol:
$$
A \to B: M,\sigma_A(M)
$$
where $\sigma_A(M)$ denotes A's digital signature on message $ M $
using $DAS$ digital signature scheme in which only the hash-value of
$M$ by a function $\hop$ is considered.  The following attack:
$$
A\to B: M',\sigma_A(M)
$$
can be launched successfully if the intruder first computes two
different messages $M$ and $M^{\prime}$ having the same hash value and
then can lead Alice into executing the protocol with message $M$.

\paragraph{Collisions in practise.}  MD5 Hash function is one of the most
widely used cryptographic hash functions nowadays.  It was designed in
1992 as an improvement on MD4, and its security was widely studied
since then by several authors.  The first result was a
pseudo-collision for MD5 \cite{BoerB93}. When permitting to change the
initialisation vector, another attack (free-start collision) has been
found \cite{Dobbertin96a}. Recently, a real collision involving two
1024-bits messages was found with the standard value \cite{WangFLY04}.
This first weakness was extended into a differential-like
attack~\cite{WangY05} and tools were developed~\cite{Vla1,Vla2} for
finding the collisions which work for any initialisation value and
which are quicker than methods presented in \cite{WangFLY04}. Finally,
other methods have been developed for finding new MD5
collisions~\cite{Yaj,Sas}.  The development of collision-finding
algorithms is not restricted to MD5 hash function.  Several methods
for MD4 research attack have been
developed~\cite{WangLFCY05,Dobbertin96}. In \cite{WangLFCY05} a method
to search RIPE-MD collision attacks was also developed, and in
\cite{BihamC04}, a collision on SHA-0 has been presented. Finally,
Wang \textit{et al.} have developed in \cite{WangYY05a} another method
to search for collisions for the SHA-1 hash function.

\paragraph{Goal of this paper.} This development of methods at the
cryptographic level to built collisions in a reasonable time have
until now not been taken into account in a symbolic model of
cryptographic protocols. We also note that the inherent complexity of
these attacks make them not representable in any computational model
that we are aware of.  In this paper we propose a decision procedure
to decide insecurity of cryptographic protocols when a hash function
for which collisions may be found is employed. Relying on the
result~\cite{ChevalierR05} we do not consider here other cryptographic
primitives such as public key encryption, signature or symmetric key
encryption, and assume that a protocol execution has already been
split into the views of the different equational theories. The
decidability proof presented here heavily relies on a recent
result~\cite{CR-RTA06} that permits to reduce constraint solving
problems with respect to a given intruder to constraint solving
problems for a simpler one. This result relies on a new notion of
\emph{mode}. This notion aims at exhibiting a modular structure in an
equational theory but has no simple intuitive meaning. In the case of
an exponential operator as treated in~\cite{CR-RTA06} the separation
was between an exponential symbol and the abelian group operations on
its exponents, whereas here the separation is introduced between the
application of the hash function and the functions employed by the
intruder to find collisions.

\paragraph{Outline.} We first give in Section~\ref{sec:setting} the
definitions relating to terms and equational theories. We then present
in Section~\ref{sec:reachability} our model of an attacker against a
protocol, and how we reduce the search for flaws to reachability
problems with respect to an intruder theory. In
Section~\ref{sec:intruder} we describe in detail how we model the fact
that an intruder may construct colliding messages, and how this
intruder theory can be decomposed into simpler intruder theories. We
give proof sketch of these reductions in
Section~\ref{sec:decidability} and conclude in
Section~\ref{sec:conclusion}.

\section{Formal setting\label{sec:setting}}

\subsection{Basic notions\label{subsec:setting:notions}}

We consider an infinite set of free constants \Constants{} and an
infinite set of variables \Variables.  For any  signature \call{G}
(\textit{i.e.}  sets of function symbols not in $C$ with arities) we
denote \gsig{\call{G}} (resp.  \vsig{\call{G}}) the set of terms
over $\call{G}\cup{}\Constants{}\change{}$ (resp.
$\call{G}\cup{}\Constants{}\change{}\cup\Variables$).  The former is
called the set of ground terms over \call{G}, while the latter is
simply called the set of terms over \call{G}.  The arity of a function
symbol $f$ is denoted by \arity{f}.  Variables are denoted by $x$,
$y$, terms are denoted by $s$, $t$, $u$, $v$, and finite sets of terms
are written $E,F,...$, and decorations thereof, respectively.  We
abbreviate $E\cup F$ by $E,F$, the union $E\cup\{t\}$ by $E,t$ and
$E\setminus \{t\}$ by $E\setminus t$.

Given a signature \call{G}, a \emph{constant} is either a free constant or
a function symbol of arity $0$ in \call{G}. We define the set of atoms
\Atoms{} to be the union of \Variables{} and the set of constants.
Given a term $t$ we denote by \Var{t} the set of variables occurring
in $t$ and by \Cons{t} the set of constants occurring in $t$.  We
denote by \atoms{t} the set \Var{t} $\cup$ \Cons{t}.  A substitution
$\sigma$ is an involutive mapping from \Variables{} to \vsig{\call{G}}
such that $\Supp{\sigma}=\{ x|\sigma(x)\not=x\}$, the \emph{support}
of $\sigma$, is a finite set.  The application of a substitution
$\sigma$ to a term $t$ (resp.  a set of terms $E$) is denoted
$t\sigma$ (resp. $E\sigma$) and is equal to the term $t$ (resp. $E$)
where all variables $x$ have been replaced by the term $\sigma(x)$. A
substitution $\sigma $ is \emph{ground} w.r.t.  $\call{G}$ if the
image of $\Supp{\sigma }$ is included in $\gsig{\call{G}}$.

An \emph{equational presentation} $\call{H}=(\call{G},A)$ is defined
by a set $A$ of equations $u=v$ with $u,v\in\TGX{}$ and $u,v$ without
free constants. For any equational presentation \call{H} the relation
$=_{\call{H}}$ denotes the equational theory generated by
$(\call{G},A)$ on \TGX{}, that is the smallest congruence containing
all instances of axioms of $A$. Abusively we shall not distinguish
between an equational presentation \call{H} over a signature \call{G}
and a set $A$ of equations presenting it and we denote both by
\call{H}. We will also often refer to \call{H} as an equational theory
(meaning the equational theory presented by \call{H}). An equational
theory \call{H} is said to be \emph{consistent} if two free constants
are not equal modulo \call{H} or, equivalently, if it has a model with
more than one element modulo \call{H}. An equational theory \call{H} is said to be \emph{regular} if for all 
equations $u=v\in A$, we have $\Var{u} =\Var{v}$.

For all signature \call{G} that we consider, we assume that
$<_\call{G}$ is a total simplification ordering on \gsig{\call{G}} for
which the minimal element is a free constant \minic{}.
\emph{Unfailing completion} permits, given an equational theory
\call{H} defined by a set $A$ of equations, to build from $A$ a
(possibly infinite) set $R(A)$ of equations $l=r$ such that the
ordered rewriting relation between terms defined by $t\to_{R(A)}t'$
if:
\begin{itemize}
\item There exists $l=r\in{}R(A)$ and a ground substitution $\sigma$
  such that $l\sigma=s$ and $r\sigma=s'$, $t=t[s]$ and
  $t'=t[s\leftarrow s']$;
\item We have $t'<_\call{G}t$.
\end{itemize}
This ordered rewriting relation is convergent, that is for all terms
$t$, all ordered rewriting sequences starting from $t$ are finite, and
they all have the same limit, called the \emph{normal form} of $t$.
We denote this term $\norm{t}_{R(A)}$, or \norm{t} when the equational
theory considered is clear from the context. In the sequel we denote
\speC{} the set consisting of \minic{} and of all symbols in \call{G}
of arity $0$.

The \emph{syntactic subterms} of a term $t$ are denoted \syncsub{t}
and are defined recursively as follows. If $t$ is an atom then
$\syncsub{t}=\set{t}$. If $t=f(t_1,\ldots,t_n)$ then
$\syncsub{t}=\set{t}\cup\bigcup_{i=1}^n\syncsub{t_i}$. The
\emph{positions} in a term $t$ are sequences of integers defined
recursively as follows, $\varepsilon$ being the empty sequence.  The
term $t$ is at position $\varepsilon$ in $t$. We also say that
$\varepsilon$ is the root position.  We write $p\leq q $ to denote
that the position $p$ is a prefix of position q.  If $u$ is a
syntactic subterm of $t$ at position $p$ and if $u=f(u_1,\ldots,u_n)$
then $u_i$ is at position $p\cdot{}i$ in $t$ for
$i\in\set{1,\ldots,n}$.  We write $t_{|p}$ the subterm of $t$ at
position $p$.  We denote $t[s]$ a term $t$ that admits $s$ as
syntactic subterm.  We denote by $\topsym{\_}$ the function that
associates to each term $t$ its root symbol.

\subsection{Mode in an equational theory}

We recall here the notion of \emph{mode} on a signature, which is
defined in~\cite{CR-RTA06}. Assume \call{H} is an equational theory
over a signature \call{G}, and let $\call{G}_0$ be a subset of
\call{G}. Assume also that the set of variables is partitioned into
two sets $\Variables_0$ and $\Variables_1$.  We first define a
signature function \sig{\_} on $\call{G}\cup\Atoms$ in the following
way:
$$
\begin{array}{rcl}
  \sig{\_} & : & \call{G}\cup \Atoms  \to  \{0,1,2\} \\
  \sig{f} &=& \left\lbrace
    \begin{array}{l}
      0 \mbox{ if } f \in \call{G}_0 \cup \call{X}_0 \\
      1 \mbox{ if } f \in (\call{G}\setminus\call{G}_0) \cup \call{X}_1\\
      2 \mbox{ otherwise,  i.e. when } f \mbox{  is a free constant} 
    \end{array}
  \right.
\end{array}
$$
The function \sig{\_} is extended to terms by taking \sig{t}=
\sig{\topsym{t}}. 

We also assume that there exists a \emph{mode} function
\mode{\cdot}{\cdot} such that \mode{f}{i} is defined for every symbol
$f\in\call{G}$ and every integer $i$ such that $1\leq i\leq
\arity{f}$.  For all valid $f,i$ we have $\mode{f}{i}\in \{0,1\} $ and
$\mode{f}{i}\le{}\sig{f}$. Thus for all $f\in \call{G}_0$ and for all
$i$ we have $\mode{f}{i}=0$.

\subsubsection{Well-moded equational theories.}

A position different from $\varepsilon $ in a term $t$ is
\emph{well-moded} if it can be written $p\cdot i$ (where $p$ is a
position and $i$ a nonnegative integer) such that $\sig{t_{|p\cdot i}}
= \mode{\topsym{t_{|p}}}{i}$.  In other words the position in a term
is well-moded if the subterm at that position is of the expected type
w.r.t. the function symbol immediately above it.  A term is {\em well-moded}
if all its \emph{non root} positions are well-moded.  Note in
particular that a well-moded term does not contain free constants. If
a position of $t$ is not well-moded we say it is \emph{ill-moded} in
$t$. A term is \emph{pure} if its only ill-moded subterms are atoms.
An equational presentation $\call{H}=(\call{G},A)$ is well-moded if
for all equations $u=v$ in $A$ the terms $u$ and $v$ are well-moded
and \sig{u}=\sig{v}. One can prove that if an equational theory is
well-moded then its completion is also well-moded~\cite{CR-RTA06}.

Note that if \call{H} is the union of two equational theories
$\call{H}_0$ and $\call{H}_1$ over two disjoint signatures
$\call{G}_0$ and $\call{G}_1$, the theory \call{H} is well-moded when
assigning mode $i$ to each argument of each operator $g\in\call{G}_i$,
for $i\in\set{0,1}$.

\subsubsection{Subterm values.}

The notion of mode also permits to define a new subterm relation in
\vsig{\call{G}}.

We call a \emph{subterm value} of a term $t$ a syntactic subterm of
$t$ that is either atomic or occurs at an ill-moded position of
$t$\footnote{Note that the root position of a term is \emph{always}
  ill-moded.}.  We denote \Sub{t} the set of subterm values of $t$.
By extension, for a set of terms $E$, the set $\Sub{E}$ is defined as
the union of the subterm values of the elements of $E$.  The subset of
the maximal and strict subterm values of a term $t$ plays an important
role in the sequel. We call these subterm values the \emph{factors} of
$t$, and denote this set \factor{t}.
\begin{example}{\label{example:subterm}}
  Consider two binary symbols $f$ and $g$ with $\sig f=\sig g=\mode f
  1 = \mode g 1 = 1$ and $\mode f 2 = \mode g 2 = 0$, and
  $t=f(f(g(a,b),f(c,c)),d)$. Its subterm values are $a$, $b$,
  $f(c,c)$, $c$, $d$, and its factors are $a$, $b$, $f(c,c)$ and $d$.
\end{example}
In the rest of this paper and unless otherwise indicated, \emph{the
  notion of subterm will refer to subterm values}. 

\subsubsection{Unification systems.}

We review here properties of well-moded theories with respect to
unification that are addressed in~\cite{CR-RTA06}. Assume \call{H} is a well-moded equational theory over a signature
\call{G}, and let $\call{H}_0$ be its projection over the signature
$\call{G}_0$ of symbols of signature $0$. Let us first define
unification systems with ordering constraints.

\begin{definition}{\label{def:unification}(Unification systems)}
  Let $\call{H}$ be a set of equational axioms on \TGX.  An$\sigma\models (\call{C}_\alpha,E \rhd{}_{\call{S}_0}m,\call{C}_\beta,\call{S})$.
  $\call{H}$-\emph{unification system} \call{S} is a finite set of
  couples of terms in \TGX{} denoted by $\{u_i \unif{}
  v_i\}_{i\in\set{1,\ldots,n}}$.  It is satisfied by a ground
  substitution $\sigma$, and we note $\sigma\models{}_{\call
    H}\call{S}$, if for all $i\in\set{1,\ldots,n}$ we have $u_i\sigma
  =_\call{H} v_i\sigma$.
\end{definition}

We will consider only satisfiability of unification systems with
ordering constraints. That is, we consider the following decision
problem:

\begin{decisionproblem}{Ordered Unifiability}
  \entreeu{A \call{H}-unification system \call{S} and an ordering
    $\prec$ on the variables $X$ and constants $C$ of \call{S}.} %
  \sortie{\textsc{Sat} iff there exists a substitution $\sigma$ such
    that $\sigma\models_{\call H}\call{S}$ and for all $x \in X$ and $
    c \in C$, $x \prec c$ implies $c \notin \syncsub{x\sigma} $ }
\end{decisionproblem}

\section{Analysis of reachability properties of cryptographic
  protocols}\label{sec:reachability}

We recall in this section the definitions of~\cite{ChevalierR05}
concerning our model of an intruder attacking actively a protocol, and
of the simultaneous constraint satisfaction problems employed to model
a finite execution of a protocol.

\subsection{Intruder deduction systems}

We first give here the general definition of intruder systems, as is
given in~\cite{ChevalierR05}. We then give the definition of a
\emph{well-moded intruder} that we will use in this paper.  In the
context of a security protocol (see \textit{e.g.}~\cite{meadows96} for
a brief overview), we model messages as ground terms and intruder
deduction rules as rewrite rules on sets of messages representing the
knowledge of an intruder.  The intruder derives new messages from a
given (finite) set of messages by applying intruder rules.  Since we
assume some equational axioms $\call{H}$ are satisfied by the function
symbols in the signature, all these derivations have to be considered
\emph{modulo} the equational congruence $=_{\call{H}}$ generated by
these axioms.  In our setting an intruder deduction rule is specified
by a term $t$ in some signature \call{G}.  Given values for the
variables of $t$ the intruder is able to generate the corresponding
instance of $t$.

\begin{definition}{\label{def:intruder}}
  An \emph{intruder system} $\call I$ is given by a triple
  \intrus{\call{G}}{\call{S}}{\call{H}} where $\call{G}$ is a
  signature, $S \subseteq \TGX$ and $\call H$ is a set of equations
  between terms in $\TGX $.  To each $t\in S$ we associate a
  \emph{deduction rule} $\lrule{t}:\Var{t}\to{}t$ and \glrule{t}
  denotes the set of ground instances of the rule \lrule{t}
  \emph{modulo \call{H}}:
  $$
  \glrule{t}=\condset{l\to{}r}{\exists\sigma, \mbox{ground
      substitution on } \call{G},~l=\Var{t}\sigma
    ~\mathrm{and}~r=_\call{H} t\sigma }
  $$
  The set of rules $\grule{}_{\call I}$ is defined as the union of the
  sets \glrule{t} for all $t\in \call{S}$.
\end{definition}

Each rule $l\to{}r$ in $\grule{}_{\call I}$ defines an intruder
deduction relation $\to_{l\to{}r}$ between finite sets of terms. Given
two finite sets of terms $E$ and $F$ we define $E\to_{l\to{}r}F$ if
and only if $l\subseteq{}E$ and $F = E \cup \set{r}$.  We denote
$\to_{\call I}$ the union of the relations $\to_{l\to{}r}$ for all
$l\to{}r$ in $L_{\call I}$ and by $\to_{\call I}^*$ the transitive
closure of $\to_{\call I}$.  Note that by definition, given sets of
terms $E$, $E'$ ,$F$ and $F'$ such that $E=_\call{H}E'$ and
$F=_\call{H}F'$ we have $E\to_{\call I}F$ iff $E'\to_{\call I}F'$. We
simply denote by $\to$ the relation $\to_{\call I}$ when there is no
ambiguity about ${\call I}$.

A \emph{derivation} $D$ of length $n$, $n\ge 0$, is a sequence of
steps of the form $E_0\to_{\call I} E_0,t_1\to_{\call
  I}\cdots\to_{\call I} E_n$ with finite sets of ground terms
$E_0,\ldots{}E_n$, and ground terms $t_1,\ldots,t_n$, such that
$E_i=E_{i-1}\cup{}\{t_i\}$ for every $i\in \set{1,\ldots,n}$.  The
term $t_n$ is called the {\em goal} of the derivation. We define
$\rhclos{E}{\call I}$ to be equal to the set $\condset{ t}{ \exists F
  \mbox{ s.t. }  E \to^*_{\call I} F \mbox{ and } t\in F} $
\textit{i.e.} the set of terms that can be derived from $E$. If there
is no ambiguity on the deduction system $\call I$ we write \clos{E}
instead of $\rhclos{E}{\call I}$.

We now define well-moded intruder systems and their properties.

\begin{definition}{\label{def:well:moded:intruder}}
  Given a well-moded equational theory \call{H}, an intruder system
  $\call{I}=\intrus{\call{G}}{S}{\call{H}}$ is \emph{well-moded} if
  all terms in $S$ are well-moded.
\end{definition}

\subsection{Simultaneous constraint satisfaction problems}

We introduce now the constraint systems to be solved for checking
protocols. It is presented in~\cite{ChevalierR05} how these constraint
systems permit to express the reachability of a state in a protocol
execution.

\begin{definition}{\label{def:constraints}(Constraint systems)} 
  Let ${\call I} =\langle \call{G}, S, \call{H} \rangle $ be an
  intruder system.  An $\call I$-\emph{Constraint system} \call{C} is
  denoted: $((E_i\rhd{}v_i)_{i\in\set{1,\ldots,n}},\call{S})$ and it
  is defined by a sequence of couples $(E_i,
  v_i)_{i\in\set{1,\ldots,n}}$ with $v_i\in\Variables{}$ and $E_i
  \subseteq \TGX$ for $i\in\set{1,\ldots,n}$, and $E_{i-1}\subseteq
  E_{i}$ for $i\in\set{2,\ldots,n}$ and by an $\call H$-unification
  system \call{S}.
  
  An $\call I$-\emph{Constraint system} \call{C} is satisfied by a
  ground substitution $\sigma$ if for all $i\in\set{1,\ldots,n}$ we
  have $v_i\sigma\in\clos{E_i\sigma}$ and if
  $\sigma\models_\call{H}{}\call{S}$.  If a ground substitution
  $\sigma$ satisfies a constraint system \call{C} we denote it by
  $\sigma\models_{\call I}\call{C}$.
\end{definition}

Constraint systems are denoted by \call{C} and decorations thereof.
Note that if a substitution $\sigma$ is a solution of a constraint
system \call{C}, by definition of constraint and unification systems
the substitution \norm{\sigma} is also a solution of \call{C}.  In the
context of cryptographic protocols the inclusion $E_{i-1}\subseteq
E_{i}$ means that the knowledge of an intruder does not decrease as
the protocol progresses: after receiving a message a honest agent will
respond to it. This response can be added to the knowledge of an
intruder who listens to all communications.

We are not interested in general constraint systems but only in those
related to protocols. In particular we need to express that a message
to be sent at some step $i$ should be built from previously received
messages recorded in the variables $v_j, j<i $, and from the initial
knowledge.  To this end we define:

\begin{definition}{\label{def:deterministic}(Deterministic Constraint Systems)}
  We say that an \call{I}-constraint system $((E_i \rhd{}
  v_i)_{i\in\set{1,\ldots,n}} , \call{S})$ is \emph{deterministic} if
  for all $i$ in $\set{1,\ldots,n}$ we have
  $\Var{E_i}\subseteq\set{v_1,\ldots,v_{i-1}}$
\end{definition}

In order to be able to combine solutions of constraints for the
intruder theory \call{I} with solutions of constraint systems for
intruders defined on a disjoint signature we have, as for unification,
to introduce some ordering constraints to be satisfied by the solution
(see~\cite{ChevalierR05} for details on this construction).
Intuitively, these ordering constraints prevent from introducing cycle
when building a global solution.  This motivates us to define the
\emph{Ordered Satisfiability} problem:

\begin{decisionproblem}{Ordered Satisfiability}
  \entreeu{an \call{I}-constraint system \call{C}, $X=\Var{\call{C}}$,
    $C=\Const{\call{C}}$ and a linear ordering $\prec$ on $X \cup C$.}
  \sortie{\textsc{Sat} iff there exists a substitution $\sigma$ such
    that $\sigma\models_{\call I}\call{C}$ and\\
    for all $x \in X$ and $ c \in C$, $x \prec c$ implies $c \notin
    \syncsub{x\sigma} $ }
\end{decisionproblem}

\section{Model of a collision-aware intruder\label{sec:intruder}}

We define in this section intruder systems to model the way an active
intruder may deliberately create collisions for the application of
hash functions. Note that our model doesn't take into account the time
for finding collisions, which is significantly greater than the time
necessary for other operations. The results that we can obtain can
therefore be seen as worst-case results, and should be assessed with
respect to the possible time deadline in the actual specification of a
protocol under analysis. Further works will also be concerned with the
fact that given a bound on intruder's deduction capabilities, a
collision may be found only with a probability p, $0\leq p\leq 1$.

We consider in this paper five different intruder models. We will
reduce in two steps the most complex one to a simpler one, relying on
the notion of well-moded theories and on the results
in~\cite{CR-RTA06}. We then prove decidability of ordered reachability
for this simpler intruder system.

\subsection{Intruder on words\label{subsec:intruder:AU}}

We first define our goal intruder, that is an intruder only able to
concatenate messages and extract \emph{prefixes} and \emph{suffixes}.
We denote $\IAU=\nintrus{AU}$ an intruder system that operates on
words, such that, if $\_\cdot\_$ denotes the concatenation and
$\epsilon$ denotes the empty word, the intruder has at its disposal
all ground instances of the following deduction rules:
$$
\left\lbrace
\begin{array}{ll}
  \deduce{x,y}{x\cdot y}\\
  \deduce{x\cdot y}{x}\\
  \deduce{x\cdot y}{y}\\
  \deduce{}{\epsilon}\\
\end{array}
\right.
$$
We moreover assume that the concatenation and empty word operations
satisfy the following equations:
$$
\left\lbrace
  \begin{array}{rclrcl}
    x\cdot (y\cdot z) &=& (x\cdot y)\cdot z&&&\\
    x\cdot \epsilon &=& x & \epsilon\cdot x &=& x\\ 
  \end{array}
\right.
$$

Given these definitions, we can see terms over \vsig{\FAU} as
\emph{words} over the alphabet $\Variables\cup\Constants$, and we
denote \letters{w} the set of atoms (either variable or free
constants) occurring in $w$. As usual, we extend \letters{\_} to set
of terms in \vsig{\FAU} by taking the union of letters occurring in
each term.

\paragraph{Pitfall.}
Notice that this intruder model does not fit into the intruder systems
definition of~\cite{ChevalierR05,CR-RTA06}. The rationale for this is
that, in the notation given here, the application of the rules is
non-deterministic, and thus cannot be modelled easily into our
``deduction by normalisation'' model. We however believe that a
deterministic and still associative model of message concatenation by
means of an ``element'' unary operator, associative operator
``$\cdot$'', and $\mathop{\mathrm{Head}}$ and $\mathop{\mathrm{Tail}}$
operations may be introduced. This means that we also assume that
unification problems are only among words of this underlying theory,
disregarding equations that may involve these extra operators. Another direction would be to extend the current definition of intruder systems to take these deductions directly into account. We leave the exact soundness of our model for further analysis and
concentrate on the treatment of collisions discovery for hash
functions.

\subsection{Intruder on words with free function
  symbols\label{subsec:intruder:free}}

We extend the \IAU{} intruder with two free function symbols 
\gop{} and \fop{}.
We first define an intruder able to compose messages using
a free function symbol \gop{} of arity $4$.
We denote  $\IG=\intrus{\set{\gop}}{\set{\gfunc xy}}{\emptyset}$ this intruder.
It has at its disposal all ground instances of the following rule:
$$\deduce{x_1,x_2,y_1,y_2}{\gfunc{x}{y}}$$
We define a similar intruder with function symbol \fop{}.
We denote $\IF=\intrus{\set{\fop}}{\set{\ffunc xy}}{\emptyset}$
this  intruder which has at its disposal all ground instances of the following rule: $$\deduce{x_1,x_2,y_1,y_2}{\ffunc{x}{y}}$$
Finally, we define \Ifree{} intruder  as
 the disjoint union of \IAU{}, \IF{} and \IG{}, and we have:
$$
\Ifree=\intrus{\FAU\cup\set{\gop,\fop}}{\SAU\cup\set{\ffunc xy,
    \gfunc xy}}{\EAU}.
$$

\subsection{Hash-colliding intruder\label{subsec:intruder:hash}}

We consider a signature modelling the following different operations:
\begin{itemize}
\item The \emph{concatenation} of two messages, the extraction of a
  suffix or a prefix of a concatenated message and the production of
  an empty message, as in the case of the \IAU{} intruder system;
\item The application of a hash function $\hop$ for which it is
  possible to find collisions, the hash-value of a message $m$ denoted
  $\hfunc m$;
\item Two function symbols $\fop$ and $\gop$ denoting the (complex)
  algorithm being used to find collisions starting from two different
  messages $m$ and $m'$.
\end{itemize}

We assume that the algorithm employed by the intruder to find
collisions starting from two messages $m$ and $m'$ proceeds as
follows:
\begin{enumerate}
\item First the intruder splits both messages into two parts, thus
  choosing $m_1,m_2,m_1',m_2'$ such that $m=m_1\cdot{}m_2$ and
  $m'=m_1'\cdot{}m_2'$;
\item Then, in order to find collisions, the intruder computes two
  messages $\gfunc m {m'}$ and $\ffunc m {m'}$ such that:
  \begin{center}
    \raisebox{-0.59ex}{(HC)}\parbox[c]{0.90\linewidth}{
      $$
      \hfunc{m_1\cdot{}\gfunc m {m'}\cdot{}m_2}=\hfunc{m_1'\cdot{}\ffunc
        m {m'}\cdot{}m_2'}
      $$
    }
  \end{center}
\end{enumerate}

A consequence of our model is that in order to build collisions
starting from two messages $m$ and $m'$ the intruder must know
(\textit{i.e.} have in its knowledge set) these two messages. A side
effect is that it is not possible to build three (or more) different
messages with the same hash value by iterating the research for
collisions. Formally, the core of the proof of this assertion is lemma~\ref{lemma:00:lem8}.

In a more comprehensive model we might moreover want to model that
collisions cannot always be found using attacks published in the
literature, but instead that given a deadline, the probability $p$ of
success of an attack is strictly below 1. This would imply that the
application of this rule by the intruder would, assuming independence
of collision attacks, reduce the likelihood of the symbolic attack
found. In this setting our model would account for attacks with a
non-negligible probability of success as is shown in~\cite{B04sasyft}.

Leaving probabilities aside, we express intruder's deductions in our
setting by adding the rule $\deduce x {\hfunc x}$ to the deduction
rules of the \Ifree{} intruder.  As a consequence, the previous
description of the \Ifree{} intruder enables us to model a
collision-capable intruder

\vspace*{1ex}

\begin{center}
\begin{tabular}[c]{rl}
  \multicolumn{2}{c}{\IH{}=\intrus{\FH}{\SH}{\EH}}\\[1em]
  \text{with:}& \hspace*{-4em}\hbox{\vbox{$\left\lbrace
        \begin{array}[c]{rcl}
          \FH&=&\FAU\cup\set{\fop,\gop,\hop}\\
          \SH&=&\SAU\cup\set{\fop(x_1,x_2,y_1,y_2),\gop(x_1,x_2,y_1,y_2),\hfunc
            x}\\
          \EH&=&\EAU \cup\set{(HC)}\\
        \end{array}
      \right.$}}\hspace*{-6em}\\
\end{tabular}
\end{center}

\vspace*{1ex}

For the following mode and signature functions the theory
$\EAU\cup\set{ (HC) }$ is a well-moded theory.
\begin{center}
  \begin{tabular}[c]{rl}
    \text{mode:}&\hspace*{-4em}\hbox{\vbox{$\left\lbrace
          \begin{array}[c]{lr}
            \mode{\cdot}{1} = \mode{\cdot}{2}= \mode{\gop{}}{i}=\mode{\fop{}}{i} = 0
            & \forall i \in\set{1,\ldots,4}\\
            \mode{\hop{}}{1} = 0& \\
          \end{array}
        \right.
        $}}\hspace*{-6em}\\
    \text{Signature:}&\hspace*{-6em}\hbox{\vbox{$\left\lbrace
          \begin{array}[c]{l}
            \sig{\cdot}=\sig{\epsilon}=\sig{\fop{}}=\sig{\gop{}}=0\\
            \sig{\hop{}}=1\\
          \end{array}
        \right.
        $}}\hspace*{-6em}\\
  \end{tabular}
\end{center}
Notice that in this case, every well-moded syntactic subterm of a term $t$ is  of signature $0$, and that every ill-moded strict syntactic subterm is of signature $1$ (lemma~\ref{lemma:00:normal}). 
The main result of this paper is the following decidability result.

\begin{theo}
  Ordered satisfiability for the \IH{} intruder is decidable.
\end{theo}

\piccaption{Reduction strategy}
\parpic[r]{
    {\large~~\xymatrix{ %
        & {\IH{} \ar[d]|{\text{\normalsize Algorithm~1}}}&\\
        & {\Ifree\ar[d]|{\text{\normalsize Generic combination algorithm~\cite{ChevalierR05}}} \ar[dl]|\hole \ar[dr]|\hole }&\\
        {\IG } & {\IF }& {\IAU{}}\\
      }~~}
  }\label{fig:red} 
  The rest of this paper is dedicated to the proof of this theorem.
  The technique employed consists in successive reductions to simpler
  problems and in finally proving that all simpler problems are
  decidable. These reductions are summarised in
  Figure $1$.  A proof for the decidability of
  the \IG{}, \IF{} and \IAU{} is given in
  Section~\ref{subsec:decidability:Ifree}.  Algorithm~1, that permits
  the first reduction, is based on the facts that the \IH{} intruder
  is well-moded (as seen above) and that we can apply a reduction
  according to the criterion of~\cite{CR-RTA06} for well-moded
  intruder systems.

  \boite{CRITERION: If $ E \to_{{\call S}_1} E,r \to_{{\call S}_1}
    E,r,t $ and $r\notin\Sub{E,t}\cup\speC$ then there is a set of
    terms $F$ such that $E \to^*_{{\call S}_0} F \to_{{\call S}_1} F,t
    $.  }

  If a well-moded intruder system system satisfies this criterion,
  then the following proposition holds. It is a cornerstone for the
  proof of completeness of Algorithm~1.

\begin{proposition}{\label{prop:shadok}}
  Let \call{I} be a well-moded intruder that satisfies the criterion,
  and let \call{C} be a deterministic \call{I}-constraint system.  If
  \call{C} is satisfiable, there exists a substitution $\sigma$ such
  that $\sigma\models_{\call{I}}\call{C}$ and:
  $$
  \set{t\in\Sub{\norm{\Sub{\call{C}}\sigma}}\vert\sig{t}=1}\subseteq\set{\norm{t\sigma}\vert ~ (t\in\Sub{\call{C}} ~ and ~ \sig{t}=1) ~ or ~ t\in\Variables}$$
\end{proposition}

\section{Decidability of reachability\label{sec:decidability}}

We present here a decision procedure for {\em Ordered Satisfiability
  Problem} for \IH{} intruder system.  Our technique consists in
simplifying the intruder system \IH{} to \Ifree{}. We then reduce the
decidability problems of ordered reachability for deterministic
constraint problems for \Ifree{} to the decidability problems of
ordered reachability for deterministic constraint problems for \IG{},
\IF{} and \IAU{}.  We finally prove the decidability for these
intruder systems.

\subsection{Reduction to \Ifree{}-intruder\label{subsec:decidability:reduction}}

\subsubsection{Algorithm}
We present here a procedure for reducing \IH{} intruder system to
\Ifree{} intruder system that takes as input a deterministic
constraint system \call{C} $=
((E_i\rhd{}v_i)_{i\in\set{1,\ldots,n}},\call{S})$ and a linear
ordering $\prec_i$ on atoms of \call{C}. Let $m=|\Sub{\call{C}}|$ be
the number of subterms in \call{C}.

\begin{itemize}
\item[]\hspace*{-1.5em}\textbf{Algorithm 1}
\item[\emph{Step 1.}] Choose a number $k\le{}m$ and add $k$ equations
  $h_j\unif{}\hfunc{c_j}$ to \call{S} where the $h_j,c_j$ are new
  variables.
\item[\emph{Step 2.}] For each
  $t\in\Sub{\call{C}}\cup\set{c_1,\ldots,c_k}$ choose a \emph{type}
  $0$, $1$ or $2$. If $t$ is of type $1$, choose $j_t\in\set{1,\ldots,k}$
  and add an equation $t\unif{}h_{j_t}$ to \call{S}.
\item[\emph{Step 3.}] For all $t,t'\in\Sub{\call{C}}$, if there exists
  $h\in\set{h_1,\ldots,h_k}$ such that $t\unif{}h$ and $t'\unif{}h$
  are in \call{S}, add to \call{S} an equation $t\unif{}t'$ to
  \call{S}.
\item[\emph{Step 4.}] Choose a subset $H$ of $\set{c_1,\ldots,c_k}\cup\set{h_1,\ldots,h_k}$ and
  guess a total order $<_d$ on $L=H\cup\set{v_1,\ldots,v_n}$ such that
  $v_i<_dv_j$ iff $i<j$. Write the obtained list $w_1,\ldots,w_{l}$.
  Let \call{S}' be the unification system obtained so far, and form:
  $\call{C}'=({(F_i\rhd{}w_i)}_{1\le{}i\le{}l},\call{S}')$ with:
  $$
  \left\lbrace
    \begin{array}{rclr}
      F_1&=&E_1&\hspace*{4cm}\\
      F_{i+1}&=&F_i\cup(E_{j+1}\setminus{}E_j)&
      \mbox{if }w_i=v_j\\
      F_{i+1}&=&F_i,w_i&\mbox{Otherwise}
    \end{array}
  \right.
  $$
\item[\emph{Step 5.}] For all  $t\in\Sub{\call{C}}$ chosen of type
  $1$, replace all occurrences of $t$ in the $F_i$ and all 
  occurrences of $t$ \emph{as a strict subterm} in $\call{S}'$ by the
  representant of its class $h_{j_t}$. Let $F_i'$ be the set $F_i$
  once this abstraction has been applied
\item[\emph{Step 6.}] Non-deterministically reduce \call{S}' to a
  unification system \call{S}'' free of \hop{} symbols, and form the
  satisfiable \Ifree{} constraint system:
  $$
  \call{C}''=({(F_i'\rhd{}w_i)}_{1\le{}i\le{}l},\call{S}'')
  $$
\end{itemize}

\subsubsection{Sketch of the completeness proof.}

Assume that the initial deterministic constraint system is
satisfiable. By Proposition~\ref{prop:shadok}, there exists a bound
substitution $\sigma$ satisfying \call{C}.
\begin{itemize}
\item Let the number $k$ chosen at Step~1 be the number of subterms
  whose top symbol is \hop{} in \Sub{\norm{\Sub{\call{C}}\sigma}}. The
  $h_j$ represent the different values of the terms of signature $1$.
  In the sequel we assume that $\sigma$ is extended to the $h_j$ such
  that all $h_j\sigma$ have a different value and are of signature
  $1$.
\item In Step~2, if $\head{\norm{t\sigma}}=1$ we choose the $j$ such
  that $\norm{t\sigma}=h_j\sigma$ and add the corresponding equation
  to \call{S}.
\item In Step~3, we add equations between terms whose normal form by
  $\sigma$ are equals in order to simplify the reduction to \Ifree{}.
\item Step~4 is slightly more intricate.  
  It relies on the fact that a rule in $\call{S}_1$ may only yield a
  term whose normal form by $\sigma$ is of signature $1$.
  
  The subset $H$ correspond to
  the subterms of signature $1$ of \Sub{\norm{\Sub{\call{C}\sigma}}}
  that are deduced by the intruder using a rule in $\call{S}_1$. 
  We then anticipate the construction of $h_i\sigma$ with the
  application of a rule in $\call{S}_1$ by requiring that the
  corresponding $c_i\sigma$ has to be build just before (lemma~\ref{lemma:00:hash2}). 
  Given the bound on $k$, this means that all remaining deductions
  performed by the intruder are now instances of rules in
  $\call{S}_0$. Since \call{C} is satisfied by $\sigma$ there exists a
  choice corresponding to \emph{quasi well-formed} derivations such
  that all remaining reachability constraints are satisfiable by
  instances of rules in $\call{S}_0$.
\item At Step~5 we ``purify'' almost all the constraint system by
  removing all occurrences of a symbol \hop{} 
  but the ones that are on the top of an equality. By the choice of
  the equivalence classes it is clear that this purification does not
  loose the satisfiability by the substitution $\sigma$.
\item The non-deterministic reduction is performed by 
  guessing whether the equality of two hashes is the consequence of a
  collision set up by the intruder or of the equality of the hashed
  messages, and will produce a constraint system \call{C}'' without
  \hop{} symbol and also satisfiable by $\sigma$ (lemma~\ref{lemma:00:reduce}).
\end{itemize}

\paragraph{Justification.}
We now justify the completeness of the algorithm with the following
lemmas.

\begin{lemma}{\label{lemma:00:consofvar}}
Let $R(\EH)$ be the completion of $\EH$ intruder theory, and let $l=r\in R(\EH)$. If $l\in\Variables$ then $l\in\Var{r}$.
\end{lemma}
\begin{proof}
Let $l=r\in R(\EH)$.
and suppose that $l\in\Variables$ and $l\notin\Var{r}$.
Let $t_1$ and $t_2$ be two different terms in $\vsig{\call{\FH}}$ and
let $\sigma_1$ and $\sigma_2$ be two substitutions such that $\sigma_1(l)=t_1, ~ \sigma_2(l)=t_2$ and $\sigma_1(r)=\sigma_2(r)$.
Then, $t_1=_{\EH}t_2$.
We deduce that if $l\in\Variables$ and $l\notin\Var{r}$ for a rule $l=r\in R(\EH)$, all terms in $\vsig{\call{\FH}}$ are equals modulo $\EH$  which is impossible.
Then for any rule $l=r\in R(\EH)$, if $l\in\Variables$, we have $l\in\Var{r}$. 
\end{proof}

\begin{lemma}{\label{lemma:00:icalp}}
Let $t$ and $t'$ be two terms in $\vsig{\call{\FH}}$.
If $t\to_{l\to{}r}t'$ and  $l\to{}r\in\grule{}_{\call I_h}$ then $l\notin\Variables$.
\end{lemma}
\begin{proof}
see proof in ~\cite{ChevalierR05}.
\end{proof}

\begin{lemma}{\label{lemma:00:normal}} 
  Let $t\in\vsig{\call{\FH}}$, we have:
  \begin{itemize}
    \item If $ t'  \in  \syncsub{t} $ and $\sig{t'}=1$ then $t'\in\Sub{t}$;
    \item If $\sig{t}=1$ then $\sig{\norm{t}}=1$. 
  \end{itemize}
\end{lemma}
\begin{proof}
$1)$ 
Let $t\in\vsig{\call{\FH}}$ and $t'\in \syncsub{t}$ such that $\sig{t'}=1$, let us prove that $t'\in\Sub{t}$. Since $t'\in\syncsub{t}$, we have two cases:
\begin{itemize}
\item
$t'=t$, then $t'\in\Sub{t}$.
\item
$t'$ is a strict syntactic subterm of $t$, then there exists an integer $p\geq 0$, an integer $i\geq 1$ such that $t_{|p.i}=t'$. 
We have $\sig{t_{|p.i}}=1$ and by definition of $\IH{}$ theory, $\mode{\topsym{t_{|p}}}{i} = 0$ then $\mode{\topsym{t_{|p}}}{i} \not= \sig{t_{|p.i}}$. Thus $t'$ is in ill-moded position in $t$, which implies that $t'\in\Sub{t}$. 
\end{itemize}

$2)$
Let $t$ be a ground term in $\gsig{\call{\FH}}$ such that $\sig{t}=1$.
We have a finite sequence of rewritings starting from $t$  leading to $\norm{t}$: $t\to_{R(\EH)}...\to_{R(\EH)}t_i\to_{R(\EH)}t_{i+1}\to_{R(\EH)}...\to_{R(\EH)}\norm{t}$.
Suppose that $\sig{t_i}=1$, and let us prove that $\sig{t_{i+1}}=1$.
Let $l=r$ be the rule applied in the step $i$. By definition of rewriting, there exists a ground substitution $\sigma$, a position $p$ such that ${t_i}_{|p}=l\sigma$,
$t_{i+1}=t_i[p\leftarrow r\sigma]$ and $l\sigma>r\sigma$.
We have two cases:
\begin{itemize}
\item If $p\not=\varepsilon$, then $\topsym{t_{i+1}}=\topsym{t_i}$ and thus by $\sig{t_i}=1$. We have $\sig{t_{i+1}}=1$.
\item If $p=\varepsilon$, then $t_i=l\sigma$. Since $\sig{l\sigma}=1$ and $l\sigma$ is ground, we have $\topsym{l\sigma}=\hop$. 
Since $l\sigma>r\sigma$ and by lemma~\ref{lemma:00:consofvar}, we have $l\notin\Variables$, and thus $l=\hfunc{l'}$ for some $l'\in\vsig{\call{\FH}}$.
Since $R(\EH)$ is well-moded and $\sig{l}=1,$ we have $\sig{r}=1.$
We have three cases:
\begin{itemize}
\item $r$ is a non-free constant.
Since the only non-free constant in $\EH$ theory is $\epsilon$ and $\sig{\epsilon}=0$, this case is impossible.
\item $r$ is a variable.
By lemma~\ref{lemma:00:consofvar}, we have $r\in\Var{l}$, and thus $r\in\syncsub{l}$. Since $l$ is well-moded in $\EH$ theory, we haven $\sig{r}=0$, which contradicts $\sig{t}=\sig{r}.$
\item
$r=\hfunc{r'}$ for $r'\in\vsig{\call{\FH}}.$
This implies that we have
$r\sigma=\hfunc{r'\sigma}$, and therefor $\sig{r\sigma}=1=\sig{t_{i+1}}$.
\end{itemize}
\end{itemize}
For all $i\in\set{1,\ldots,n-1}$, we have $\sig{t_i}=1$ implies $\sig{t_{i+1}}=1$, which proves the second point of the lemma.
\end{proof}

\begin{lemma}{\label{lemma:00:lem4}}
  Assume $E$ and $F$ are in normal form.  If $E\to_{\call S}F$ and $t
  \in \Sub{F}\setminus\set{\Sub{E}\bigcup\speC}$, Then $F \setminus E
  = t$ and $E\to_{\lrule{u}}F$, with $u\in S$ and $\sig{u}=\sig{t}$.
\end{lemma}
\begin{proof}
see proof in ~\cite{CR-RTA06}.
\end{proof}

\begin{lemma}{\label{lemma:00:consvar}}
Let $A$ and $R(A)$ be an equational theory and its completion respectively.
If $A$ is regular then $R(A)$ is regular to. 
\end{lemma}
\begin{proof}
Let $A$ be a regular equational theory, that is for all $l=r\in A$ we have $\Var{l}=\Var{r}$.
Let $l=r$ and $g=d$ be two rules such that $\Var{l}=\Var{r}$ and $\Var{g}=\Var{d}$.
Suppose that there exists a principal unifier $\sigma$ of $g$ and a non-variable subterm $l_{|p}$ of $l$.
Let us prove that the derived rule obtained by the completion algorithm $r\sigma=l\sigma[p\leftarrow d\sigma]$ preserves variables.
We have $\Var{l\sigma}=(\Var{l\sigma}\setminus\Var{l\sigma_{|p}})\cup\Var{g\sigma}$ and 
$\Var{l\sigma[p\leftarrow d\sigma]}=\Var{l\sigma}\setminus\Var{l\sigma_{|p}}\cup\Var{d\sigma},$ and since $\Var{g}=\Var{d}$, then we have $\Var{l\sigma[p\leftarrow d\sigma]}=\Var{l\sigma}=\Var{r\sigma}.$
This concludes the proof of the lemma.
\end{proof}

\begin{lemma}{\label{lemma:00:preservation}}
Let $t\in\gsig{\call{\FH}}$ with all its factors in normal form.
We have: $\Sub{t}\setminus \set{\epsilon,t} \subseteq \Sub{\norm{t}}$.
\end{lemma}
\begin{proof}
Let $t\in\gsig{\call{\FH}}$.
There exists a finite sequence of rewritings starting from $t$ leading to $\norm{t}$: 
$t\to_{R(\EH)}...\to_{R(\EH)}t_i\to_{R(\EH)}t_{i+1}\to_{R(\EH)}...\to_{R(\EH)}\norm{t}$.
Let us prove the lemma by contradiction and assume that  $u\in\Sub{t_i}\setminus\set{\epsilon,t_i}$ and $u\notin\Sub{t_{i+1}}.$
Since $u\in\Sub{t_i}\setminus \set {\epsilon ,t_i}$, there exists an integer $q\geq 1$ such that $t_i{_{|q}}=u.$
Let $l=l'$ be the rule applied on $t_i$. There exists an integer $p\geq 0$, a ground substitution $\sigma$
such that ${t_i}_{|p}=l\sigma$ and $t_{i+1}=t_i[p\leftarrow l'\sigma]$ with $l\sigma>l'\sigma.$
\begin{itemize}
\item If $u\notin\Sub{l\sigma}$ then $u\in\Sub{t_{i+1}}$.
\item If $u\in\Sub{l\sigma}$, by the fact that $l$ is well-moded, $u$ is in normal form 
and $u\not=\epsilon$, there exists $x\in\Var{l}$ such that $u\in\Sub{x\sigma}$.
Since $\Var{l}=\Var{l'}$, we have  $u\in\Sub{t_{i+1}}$. 
\end{itemize}
In the two cases, we lead to a contradiction with $u\notin\Sub{t_{i+1}}$.
This concludes the proof of the lemma.
\end{proof}

\begin{lemma}{\label{lemma:00:hyp1}}
  The intruder system \IH{} satisfies CRITERION.
\end{lemma}

\begin{proof}
Let E be a set of terms in normal forms satisfying the following derivation:
$E\to_{\call{S}_0}E,r\to_{\call{S}_1}E,r,t$ such that $r\notin\Sub{E,t}\cup\speC$.
In order to prove that there exists a set of terms F such that $E\to^*_{\call{S}_0}F\to_{\call{S}_1}F,t$, it suffices to prove that $E\to_{\call{S}_1}E,t$.
We have $E\to_{\call{S}_1}E,r$ and the only $\call{S}_1$ rule is $x\to\hfunc{x}$. By definition, there exists a normal ground substitution $\sigma$ such that 
$x\sigma \in E$
and $r=\norm{\hfunc{x\sigma}}$. Since $\sig{\hfunc{x\sigma}}=1$ by lemma~\ref{lemma:00:normal}, we have $\sig{r}=1$. Since $E,r\to_{\call{S}_1}E,r,t$, there exists a normal ground substitution $\sigma'$ such that $x\sigma'\in E,r$ and $t=\norm{\hfunc{x\sigma'}}$.
If $x\sigma'=r$, we have  $t=\norm{\hfunc{r}}$.
$\hop(r)$ is in normal form, since all its factors are in normal form and  $r\in\Sub{\hfunc{r}}\setminus\set{\hfunc{r},\epsilon}$, by lemma~\ref{lemma:00:preservation} $r\in\Sub{t},$ which contradicts the hypothesis $r\notin\Sub{E,t}\cup\speC$.
By contradiction, we have  $x\sigma'\in E$ and thus $E\to_{\call{S}_1}E,t$. 
\end{proof}

In the following lemma, $t=^1_{HC}t'$ denotes that there exists a one step rewriting between $t$ and $t'$ using (HC) rule.

\begin{lemma}{\label{lemma:00:lem8}}
Let $t_0,t,t'\in\vsig{\call{\FH}}$ such that $t_0=_{\call{\EAU}}t=^1_{HC}t'$ and $t_0=\hop(t_1\cdot{}\fop(t_1,t_2,t_3,t_4)\cdot t_2)$.
We have: $t'=_{\call{\EAU}}\hop(t_3\cdot{}\gop(t_1,t_2,t_3,t_4)\cdot{}t_4)$.
\end{lemma}

\begin{proof}
Let $\hop(m_1\cdot\fop/\gop(m_1,m_2,m_3,m_4)\cdot{}m_2)=\hop(m_3\cdot{}\gop/\fop(m_1,m_2,m_3,m_4)\cdot{}m_4)$
be the ground instance of (HC) used between $t$ and $t'$.
Let us prove that $m_1=_{\call{\EAU}}t_1$.
If $m_1\not=_{\call{\EAU}}t_1$, we have either $m_1$ is a prefix modulo $\call{\EAU}$ of $t_1$ or $t_1$ is a prefix modulo $\call{\EAU}$ of $m_1$. Let us review these two cases:
\begin{itemize}
\item
$m_1$ is a prefix modulo $\call{\EAU}$ of $t_1$:
then $t_1=m_1\cdot{}x$ and $x \not=_{\call{\EAU}} \epsilon$, then $\fop/\gop(m_1,m_2,m_3,m_4)\in\syncsub{t_1}$, then $m_2\in\syncsub{t_1}$.
And we have $m_2=y\cdot{}t_2$ with $y\not=_{\call{\EAU}}\epsilon$, then $\fop(t_1,t_2,t_3,t_4)\in\syncsub{m_2}$ then $t_1\in\syncsub{m_2}.$
We conclude that $t_1$ is a strict subterm of $m_2$ and $m_2$ is a strict subterm of $t_1$ which is impossible.
\item 
$t_1$ is a prefix modulo $\call{\EAU}$ of $m_1$: by reasoning as above on $t_2$ which is a suffix of $m_2$, we can also prove that this case is impossible.
\end{itemize}
Thus we have $m_1=_{\call{\EAU}}t_1$, and thus
$\fop/\gop(m_1,m_2,m_3,m_4)=_{\call{\EAU}}\fop(t_1,t_2,t_3,t_4)$, that is $m_i=_{\call{\EAU}}t_i$ for $i\in\set{1,2,3,4}$ and
$t'=_{\call{\EAU}}\hop(t_3\cdot{}\gop(t_1,t_2,t_3,t_4)\cdot{}t_4)$.
\end{proof}

In the following lemma, $t=^1_{\call{\EH}}t'$ denotes that there exists a finite sequence of rewritings between $t$ and $t'$ using $\call{\EAU}$ rules and where (HC) rule is used exactly one time.

\begin{lemma}{\label{lemma:00:reduce}}
  Let $\hfunc{m},\hfunc{m'}$ be two pure terms and $\sigma$ be ground
  substitution such that $\sigma\models_{\call{\EH}}\hfunc{m}\unif{}\hfunc{m'}$.
  Then either:
  $$
  \mbox{or}\left\lbrace
    \begin{array}[c]{l}
      \sigma\models_{\call{\EAU}} m\unif{}m'\\
      \sigma\models_{\call{\EAU}} \set{m\unif{}x_1\cdot{}\gfunc{x}{y}\cdot{}x_2,
        m'\unif{}y_1\cdot{}\ffunc{x}{y}\cdot{}y_2}
    \end{array}
  \right.
  $$
  with $x_1,x_2,y_1,y_2$ new variables (modulo the commutativity of
  \unif{}).
\end{lemma}
\begin{proof}
Let $m_1,m_2,m_3\in\vsig{\call{\EH}}$ such that $\hop(m_1)=^1_{HC}\hop(m_2)=^1_{HC}\hop(m_3)$.
If $m_1=_{\call{\EAU}}t_1\cdot{}\fop(t_1,t_2,t_3,t_4)\cdot{}t_2$  then, by lemma~\ref{lemma:00:lem8} we have
 $$
   \left\lbrace
    \begin{array}[c]{l}
      m_2=_{\call{\EAU}}t_3\cdot{}\gop(t_1,t_2,t_3,t_4)\cdot{}t_4\\
      m_3=_{\call{\EAU}}t_1\cdot{}\fop(t_1,t_2,t_3,t_4)\cdot{}t_2
    \end{array}
   \right.$$
Let $S_{m_1}=\set{m|~\hop(m)=_{\call{\EH}}\hop(m_1)}$
then, by lemma~\ref{lemma:00:lem8} we have 
$S_{m_1}=\set{m|~m=_{\call{\EAU}}m_1}\cup\set{m|~m=_{\call{\EAU}}t_3\cdot{}\gop(t_1,t_2,t_3,t_4)\cdot{}t_4}$.\\
We have $\sigma\models_{\call{\EH}}\hfunc{m}\unif{}\hfunc{m'}$ that is $\hfunc{m\sigma}=_{\call{\EH}}\hfunc{m'\sigma}$, and thus $m'\sigma\in S_{m\sigma}$ which implies that either $m\sigma=_{\call{\EAU}}m'\sigma$ and then 
$\sigma\models_{\call{\EAU}} m\unif{} m'$ 
or $m\sigma=_{\call{\EAU}}x_1\sigma\cdot{}\fop(x_1\sigma,x_2\sigma,y_1\sigma,y_2\sigma)\cdot{}x_2\sigma$ and $m'\sigma=_{\call{\EAU}}y_1\sigma\cdot{}\gop(x_1\sigma,x_2\sigma,y_1\sigma,y_2\sigma)\cdot{}y_2\sigma$ and then 
$\sigma\models_{\call{\EAU}} \set{m\unif{}x_1\cdot{}\gfunc{x}{y}\cdot{}x_2,
        m'\unif{}y_1\cdot{}\ffunc{x}{y}\cdot{}y_2}.$
\end{proof}

In the following lemma, we use \Ifree{} intruder with $\sig{\epsilon}=\sig{\cdot{}}=0$, $\sig{\fop}=\sig{\gop}=1$ and the notion of {\em subterms values} is defined as in ~\cite{ChevalierR05}.

\begin{lemma}{\label{lemma:00:fcons}}
Let  $E$ be a set of terms in normal form.
If $E\to^*_{\call{S}_0} \fop(t_1,t_2,t'_1,t'_2)$ and $\fop(t_1,t_2,t'_1,t'_2)\notin\syncsub{E}$ then  $E\to^*_{\call{S}_0}t_1,t_2,t'_1,t'_2$. 
\end{lemma}
\begin{proof}
We have $E\to^*_{\call{S}_0} \fop(t_1,t_2,t'_1,t'_2)$ that is, there exists a finite sequence of rewritings starting from $E$ leading to
$\fop(t_1,t_2,t'_1,t'_2)$: $E\to_{\call{S}_0}E_1\to_{\call{S}_0}...\to_{\call{S}_0}E_{n-1}\to_{\call{S}_0}E_{n-1},\fop(t_1,t_2,t'_1,t'_2)$.
By hypothesis, we have $\fop(t_1,t_2,t'_1,t'_2)\in\Sub{E_n}\setminus(\Sub{E}\cup\speC)$.
Let $E_i$ be the smallest set in the derivation such that $\fop(t_1,t_2,t'_1,t'_2)\in\Sub{E_i}\setminus(\Sub{E_{i-1}}\cup\speC)$ [$i\geq1$].
By lemma~\ref{lemma:00:lem4}, the rule applied in the step $i$ of derivation is either $x_1,x_2,y_1,y_2\to\gop(x_1,x_2,y_1,y_2)$ or $x_1,x_2,y_1,y_2\to\fop(x_1,x_2,y_1,y_2)$ 
and in our case it is $x_1,x_2,y_1,y_2\to\fop(x_1,x_2,y_1,y_2)$.
By definition, there exists a normal ground substitution $\sigma$ such that  $t_i=x_i\sigma$ and $t'_i=y_i\sigma$ for $i\in\set{1,2}$ and $t_1,t_2,t'_1,t'_2\in E_{i-1}$.
We deduce that $E\to^*_{\call{S}_0}t_1,t_2,t'_1,t'_2$.
\end{proof}

\begin{lemma}{\label{lemma:00:hash1}}
  Let \call{C} be a deterministic constraint system of the form $
  ((E_i\rhd{}v_i)_{i\in\set{1,\ldots,n}},\call{S})$ such that
  no term appearing in \call{C} has the form $\fop(t_1,t_2,t_3,t_4)$ or $\gop(t_1,t_2,t_3,t_4)$ for
  some $t_1,\ldots,t_4$, and let 
  $(h(m_1)\unif{}h(m_2))\in\call{S}$. Let $\sigma$ be a ground substitution which satisfies \call{C}.
  For all $i\in\set{1,\hdots,n}$, we have:
  $$
  \sigma\models(E_i \rhd{_{\call{S}_0}} m_1) ~\mbox{ iff }~ \sigma
  \models(E_i \rhd{_{\call{S}_0}} m_2)
  $$
\end{lemma}

\begin{proof}
  By symmetry, it suffices to prove that if $\sigma\models(E_i \rhd{_{\call{S}_0}} m_1)$ then $\sigma\models(E_i \rhd{_{\call{S}_0}} m_2)$.
  Since $\sigma\models_{\call{\EH}}(h(m_1)\unif{}h(m_2))$, by lemma~\ref{lemma:00:reduce} we have two cases:
 \begin{itemize}
 \item  If $\sigma\models_{\call{\EAU}} m_1 \unif{} m_2$ then the result is obvious.
 \item  If $\sigma\models_{\call{\EAU}} \set{m\unif{}x_1\cdot{}\gfunc{x}{y}\cdot{}x_2,
        m'\unif{}y_1\cdot{}\ffunc{x}{y}\cdot{}y_2}$ 
then  $$
   \left\lbrace
    \begin{array}[c]{l}
      m_1\sigma=_{\call{\EAU}}x_1\sigma\cdot{}\fop(x_1\sigma,x_2\sigma,y_1\sigma,y_2\sigma)\cdot{}x_2\sigma\\
      m_2\sigma=_{\call{\EAU}}y_1\sigma\cdot{}\gop(x_1\sigma,x_2\sigma,y_1\sigma,y_2\sigma)\cdot{}y_2\sigma
    \end{array}
   \right.$$
Since $\sigma\models{}(E_i\rhd{_{\call{S}_0}}m_1)$, we have $\norm{E_i\sigma}\to^*_{\call{S}_0}\norm{x_1\sigma\cdot{}\fop(x_1\sigma,x_2\sigma,y_1\sigma,y_2\sigma)\cdot{}x_2\sigma}$ and thus, 
$\norm{E_i\sigma}\to^*_{\call{S}_0}\norm{x_1\sigma}\cdot{}\fop(\norm{x_1\sigma},\norm{x_2\sigma},\norm{y_1\sigma},\norm{y_2\sigma})\cdot{}\norm{x_2\sigma}$ which implies that
$\norm{E_i\sigma}\to^*_{\call{S}_0}\norm{x_1\sigma},\norm{x_2\sigma},\fop(\norm{x_1\sigma},\norm{x_2\sigma},\norm{y_1\sigma},\norm{y_2\sigma})$.
\\
Since $\norm{E_i\sigma}\to^*_{\call{S}_0}\fop(\norm{x_1\sigma},\norm{x_2\sigma},\norm{y_1\sigma},\norm{y_2\sigma})$, we have two cases:
\begin{itemize}
\item $\fop(\norm{x_1\sigma},\norm{x_2\sigma},\norm{y_1\sigma},\norm{y_2\sigma})\notin\syncsub{\norm{E_i\sigma}}$, by lemma~\ref{lemma:00:fcons} we have  $\norm{E_i\sigma}\to^*_{\call{S}_0}\norm{x_1\sigma},\norm{x_2\sigma},\norm{y_1\sigma},\norm{y_2\sigma}$ and thus   $\norm{E_i\sigma}\to^*_{\call{S}_0}\gop(\norm{x_1\sigma},\norm{x_2\sigma},\norm{y_1\sigma},\norm{y_2\sigma})$ which implies that $\norm{E_i\sigma}\to^*_{\call{S}_0}\norm{y_1\sigma}\cdot{}\gop(\norm{x_1\sigma},\norm{x_2\sigma},\norm{y_1\sigma},\norm{y_2\sigma})\cdot{}\norm{y_2\sigma}$.
We conclude that $\sigma\models(E_i \rhd{_{\call{S}_0}} m_2)$.
\item $\fop(\norm{x_1\sigma},\norm{x_2\sigma},\norm{y_1\sigma},\norm{y_2\sigma})\in\syncsub{\norm{E_i\sigma}}$, there exists $v_j\in\Var{E_i}$ such that 
$\fop(x_1\sigma,x_2\sigma,y_1\sigma,y_2\sigma)\in\syncsub{v_j\sigma}$, with $j<i$ and $\sigma\models(E_j\rhd{_{\call{S}_0}}v_j)$.
Let $l$ be the smallest integer such that $\norm{E_l\sigma}\to^*_{\call{S}_0}\fop(\norm{x_1\sigma},\norm{x_2\sigma},\norm{y_1\sigma},\norm{y_2\sigma})$ and $\fop(\norm{x_1\sigma},\norm{x_2\sigma},\norm{y_1\sigma},\norm{y_2\sigma})\notin\syncsub{\norm{E_l\sigma}}$.
By lemma~\ref{lemma:00:fcons}, we have $\norm{E_l\sigma}\to^*_{\call{S}_0}\norm{x_1\sigma},\norm{x_2\sigma},\norm{y_1\sigma},\norm{y_2\sigma}$
and thus $\norm{E_i\sigma}\to^*_{\call{S}_0}\norm{x_1\sigma},\norm{x_2\sigma},\norm{y_1\sigma},\norm{y_2\sigma}$
which implies that $\norm{E_i\sigma}\to^*_{\call{S}_0}\gop(\norm{x_1\sigma},\norm{x_2\sigma},\norm{y_1\sigma},\norm{y_2\sigma})$.
We conclude that $\sigma\models(E_i \rhd{_{\call{S}_0}} m_2)$.
\end{itemize}
\end{itemize}
\end{proof}

\begin{lemma}{\label{lemma:00:hash2}}
  Let $\call{C}= ((E_i\rhd{}v_i)_{i\in\set{1,\ldots,n}},\call{S})$ be a deterministic constraint system such that
  no term appearing in \call{C} has the form $\fop(t_1,t_2,t_3,t_4)$ or $\gop(t_1,t_2,t_3,t_4)$ for
  some $t_1,\ldots,t_4$ and  $v_j\unif{}\hop(m)\in\call{S}$.    
  Let $\sigma$ be a ground substitution such that $\sigma\models\call{C}$
  and for all $E\rhd v \in \call{C}$, there exists a derivation starting from $\norm{E\sigma}$ leading to $\norm{v\sigma}$ where all steps use $\call{S}_0$ rules except possibly the last one which may uses $\call{S}_1$ rule. We have either
  $\sigma\models ((E_1 \rhd{}v_1,\ldots,E_j\rhd{}_{\call{S}_0}v_j,\ldots,E_n\rhd{}v_n),\call{S})$
    or
  $\sigma\models ((E_1 \rhd{}v_1,\ldots,E_j\rhd{}_{\call{S}_0}v_j',\ldots,E_n\rhd{}v_n),\call{S}')$ where
 $\call{S}'=\call{S}\cup\set{v_j'\unif{}m}$.
\end{lemma}

\begin{proof}
Let $\call{C}= ((E_i\rhd{}v_i)_{i\in\set{1,\ldots,n}},\call{S})$,  $v_j\unif{}\hop(m)\in\call{S}$ and $\sigma$ be a ground substitution such that $\sigma\models\call{C}$.
We have $\sigma\models(E_j\rhd{}v_j)$ and $v_j\sigma=_\call{\EH}\hop(m\sigma)$, that is 
there exists a finite sequence of rewritings starting from $\norm{E_j\sigma}$ leading to $\norm{\hop(m\sigma)}$ where all steps in the derivation use $\call{S}_0$ rules except possibly the last one which may uses $\call{S}_1$ rule. We have two cases:
\begin{itemize}
\item
If all used rules are of type $\call{S}_0$ then $\sigma\models E_j\rhd{}_{\call{S}_0}\hop(m)$ and thus, 
$\sigma\models ((E_1 \rhd{}v_1,\ldots,E_j\rhd{}_{\call{S}_0}v_j,\ldots,E_n\rhd{}v_n),\call{S})$.
\item 
If the last used rule is of type $\call{S}_1$ then $\norm{E_j\sigma}\to^*_{\call{S}_0}F,\norm{t\sigma}\to_{\call{S}_1}F,\norm{t\sigma},\norm{\hop(\norm{t\sigma})}$ with $\norm{\hop(\norm{t\sigma})}=\norm{\hop(m\sigma)}$ and thus, we have 
two cases for the equation $v_j\unif{}\hop(m)$,
If $\sigma\models_{\call{\EAU}} t\unif{}m$ then $\sigma\models E_j\rhd{}_{\call{S}_0} m$  
and thus we have $\sigma\models ((E_1 \rhd{}v_1,\ldots,E_j\rhd{}_{\call{S}_0}v_j',\ldots,E_n\rhd{}v_n),\call{S}')$ where
 $\call{S}'=\call{S}\cup\set{v_j'\unif{}m}$.\\
Else, the hypothesis of this lemma (no term appearing in \call{C} has the form $\fop(t_1,t_2,t_3,t_4)$ or $\gop(t_1,t_2,t_3,t_4)$ for
  some $t_1,\ldots,t_4$) permits to apply lemma~\ref{lemma:00:hash1} which implies that 
$\sigma\models E_j\rhd{}_{\call{S}_0}m$, and thus,  
$\sigma\models ((E_1 \rhd{}v_1,\ldots,E_j\rhd{}_{\call{S}_0}v_j',\ldots,E_n\rhd{}v_n),\call{S}')$ where
 $\call{S}'=\call{S}\cup\set{v_j'\unif{}m}$.
\end{itemize}
\end{proof}

\subsection{Decidability of reachability for the
  \Ifree{}-intruder \label{subsec:decidability:Ifree}}

We first reduce the \Ifree{} intruder system to simpler intruder
systems using the combination result of~\cite{ChevalierR05}. We will
consider the decidability of these subsystems in the remainder of this
section.

\begin{theo}
  Ordered satisfiability for the $\Ifree{}$ intruder
  system is decidable.
\end{theo}

 \begin{proof}
    $\Ifree{}$ intruder theory is the disjoint union of $\IAU{}$ ,
    $\IG{}$ and $\IF{}$ intruder theories.  The reachability problems
    of the three preceding theories are decidable
    (Theorem~\ref{theo:00:decidAU} and Theorem~\ref{theo:00:decidG}).
    The result obtained in ~\cite{ChevalierR05} prove that the
    disjoint union of decidable intruder theories is also decidable.
    Thus \Ifree{} is decidable.
  \end{proof}

\paragraph{Decidability of reachability for the
  \IG{}-intruder\label{subsec:decidability:G}.}

In this subsection, we consider an $\IG{}$ intruder system with
$\IG{}=\intrus{\gop}{\gop(x_1,x_2,x'_1,x'_2)}{\emptyset}$.  This
intruder has at its disposal all ground instances of the following
deduction rule:
$$
x_1,x_2,y_1,y_2\to\gop(x_1,x_2,y_1,y_2)
$$ 


\begin{theo}{\label{theo:00:decidG}}
Ordered satisfiability for $\IG{}$ intruder is decidable.
\end{theo}
\begin{proof}
Let $\call{C}$ be an $\IG{}$ deterministic constraint system.
Since $\IG{}$ intruder theory verifies the convergent public-collapsing property of ~\cite{dj-unif-2004}, $\call{S}$ contains finite equations and  $\call{C}$ contains a finite number of intruder constraints $E_i\rhd v_i$ and it is well-formed, we have ordered satisfiability problem for $\IG{}$ is decidable by the theorem 1 of~\cite{dj-unif-2004}.
\end{proof}

\paragraph{Decidability of reachability for the
  \IAU{} intruder\label{subsec:decidability:AU}.}

We now give a proof sketch for the decidability of ordered
satisfiability for the $\call{I}_{\mathrm{AU}}$ intruder.

\begin{theo}{\label{theo:00:decidAU}}
  Ordered satisfiability for the \IAU{} intruder system is decidable.
\end{theo}
\begin{proof}
  The algorithm proceeds as follows:
  \begin{itemize}
  \item Transform the deduction constraints $E\rhd{}v$ into an
    ordering constraint $<_d$;
  \item Check that $<\stackrel{\text{def}}{=}<_d \cup <_i$ is still a partial order on atoms
    of \call{C};
  \item Solve the unification problems with linear constant restriction
    $<$.
  \end{itemize}
 
  Let $\call{C}=({(E_i\rhd{}v_i)}_{0\le{}i\le{}n},\call{S})$ be a
  deterministic constraint system for the $\call{I}_{\mathrm{AU}}$
  intruder, $<_i$ be a (partial) order on
  $\Cons{\call{C}}\cup\Var{\call{C}}$, and let $\sigma$ be a solution
  of the $(\call{C},<_i)$ ordered satisfiability problem.

  Given a set of terms $E\subseteq\vsig{\FAU}$, let us denote
  $\Kappa_\call{C}=(\Cons{\call{C}}\setminus\letters{E})\setminus\Variables$.
  In plain words, $\Kappa_\call{C}(E)$ is the set of constants in
  \call{C} \textbf{not} occurring in $E$. We are now ready to define
  $<_d$ as a partial order on
  $\Cons{\call{C}}\cup\set{v_0,\ldots,v_n}$: We set $v_i<_d c$ for all
  constants $c$ in $\Kappa_\call{C}(E_i)$.

  \vspace*{+0.2em}
  \enlargethispage{1em}

  \begin{claim}
    For all $\sigma$, we have $\sigma\models(\call{C},<_i)$ if, and
    only if, $\sigma\models(\call{S},<_i\cup <_d)$
  \end{claim}

  \vspace*{+0.3em}

  \begin{proofclaim}
    Let us first prove the direct implication. Let $\sigma$ be a
    ground solution of the $(\call{C},<_i)$ ordered satisfiability
    problem.  By definition we have that $\sigma$ is a solution of
    $(\call{S},<_i)$ ordered unifiability problem. Since for all
    $0\le{}i\le{}n$ we have $\sigma\models{}E_i\rhd{}v_i$, we easily
    see that $\letters{\norm{v_i\sigma}}\subseteq\Cons{E_i}$, and
    therefore
    $\letters{\norm{v_i\sigma}}\cap\Kappa_\call{C}(E_i)=\emptyset$.
    Thus $\sigma$ is also a solution of $(\call{S},<_d\cup <_i)$.

    Conversely, assume now that $\sigma$ is a ground solution of
    $(\call{S},<_d\cup <_i)$. By definition for all $0\le{}i\le{}n$ we
    have
    $\letters{\norm{v_i\sigma}}\cap\Kappa_\call{C}(E_i)=\emptyset$,
    and thus $\letters{\norm{v_i\sigma}}\subseteq{}
    \letters{E_i}\setminus\Variables$. Thus we have
    $\norm{v_i\sigma}\in\clos{\norm{E_i\sigma}}$ for all
    $0\le{}i\le{}n$, and thus $\sigma\models(\call{C},<_i)$
  \end{proofclaim}

  Since unifiability with linear constant restriction is decidable for
  the $AU$ equational theory~\cite{Schulz90}, this finishes the proof
  of the theorem. Note that the exact complexity is not known, but the
  problem is NP-hard and solvable in
  PSPACE~\cite{Plandowski99,Plandowski04}, and it is conjectured to be
  in NP~\cite{Plandowski98,DBLP}.
\end{proof}

\section{Conclusion}
\label{sec:conclusion}

We have presented here a novel decision procedure for the search for
attacks on protocols employing hash functions subject to collision
attacks. Since this procedure is of practical interest for the
analysis of the already normalised protocols relying on these weak
functions, we plan to implement it into an already existing tool,
CL-Atse~\cite{DBLP:conf/RTA/turuani'06}. Alternatively an implementation may be done in
OFMC~\cite{DBLP:journals/ijisec/BasinMV05}, though the support of
associative operators is still partial. In order to model hash functions we have introduced
new symbols to denote the ability to create messages with the same
hash value. This introduction amounts to the skolemisation of the
equational property describing the existence of collisions. We believe
that this construction can be extended to model the more complex and
game-based properties that appear when relating a symbolic and a
concrete model of cryptographic primitives.

\bibliography{constraint}

\begin{thebibliography}{10}

\bibitem{DBLP:journals/ijisec/BasinMV05}
David~A. Basin, Sebastian M{\"o}dersheim, and Luca Vigan{\`o}.
\newblock Ofmc: A symbolic model checker for security protocols.
\newblock {\em Int. J. Inf. Sec.}, 4(3):181--208, 2005.

\bibitem{B04sasyft}
Mathieu Baudet.
\newblock Random polynomial-time attacks and {D}olev-{Y}ao models.
\newblock In Siva Anantharaman, editor, {\em {P}roceedings of the {W}orkshop on
  {S}ecurity of {S}ystems: {F}ormalism and {T}ools ({SASYFT}'04)}, Orl{\'e}ans,
  France, June 2004.

\bibitem{BihamC04}
E.~Biham and R.~Chen.
\newblock Near-collisions of sha-0.
\newblock In M.~K. Franklin, editor, {\em CRYPTO}, volume 3152 of {\em Lecture
  Notes in Computer Science}, pages 290--305. Springer, 2004.

\bibitem{ChevalierR05}
Y.~Chevalier and M.~Rusinowitch.
\newblock Combining intruder theories.
\newblock In L.~Caires, G.~F. Italiano, L.~Monteiro, C.~Palamidessi, and
  M.~Yung, editors, {\em ICALP}, volume 3580 of {\em Lecture Notes in Computer
  Science}, pages 639--651. Springer, 2005.

\bibitem{CR-RTA06}
Y.~Chevalier and M.~Rusinowitch.
\newblock Hierarchical combination of intruder theories.
\newblock In {\em {P}roceedings of the 17th {I}nternational {C}onference on
  {R}ewriting {T}echniques and {A}pplications ({RTA}'05)}, Lecture Notes in
  Computer Science, Seattle, USA, August 2006. Springer.
\newblock To appear.

\bibitem{DBLP:conf/eurocrypt/2005}
R.~Cramer, editor.
\newblock {\em Advances in Cryptology - EUROCRYPT 2005, 24th Annual
  International Conference on the Theory and Applications of Cryptographic
  Techniques, Aarhus, Denmark, May 22-26, 2005, Proceedings}, volume 3494 of
  {\em Lecture Notes in Computer Science}. Springer, 2005.

\bibitem{dj-unif-2004}
S.~Delaune and F.~Jacquemard.
\newblock Narrowing$-$based constraint solving for the verification of security
  protocols.
\newblock In {\em {P}roceedings of the 18th {I}nternational {W}orkshop of
  {U}nification ({UNIF}'04)}, Cork, Ireland, 2004.

\bibitem{BoerB93}
B.~den Boer and A.~Bosselaers.
\newblock Collisions for the compressin function of md5.
\newblock In {\em EUROCRYPT}, pages 293--304, 1993.

\bibitem{Dobbertin96}
H.~Dobbertin.
\newblock Cryptanalysis of md4.
\newblock In D.~Gollmann, editor, {\em Fast Software Encryption}, volume 1039
  of {\em Lecture Notes in Computer Science}, pages 53--69. Springer, 1996.

\bibitem{Dobbertin96a}
H.~Dobbertin.
\newblock Cryptanalysis of md5 compress.
\newblock Presented at the rumps session {\em of Eurocrypt'96}, 1996.

\bibitem{Vla2}
V.~Kl{\`\i}ma.
\newblock Finding md5 collisions - a toy for a notebook, 2005.
\newblock Cryptology ePrint Archive, Report 2005/075.
  \url{http://eprint.iacr.org/}.

\bibitem{Vla1}
V.~Kl{\`\i}ma.
\newblock Finding md5 collisions on a notebook pc using multi-mes\-sage
  modificatons, 2005.
\newblock Cryptology ePrint Archive, Report 2005/102.
  \url{http://eprint.iacr.org/}.

\bibitem{DBLP}
Kim~Guldstrand Larsen, Sven Skyum, and Glynn Winskel.
\newblock Automata, languages and programming, 25th international colloquium,
  icalp'98, aalborg, denmark, july 13-17, 1998, proceedings.
\newblock In {\em ICALP}, volume 1443 of {\em Lecture notes in Computer
  Science}. Springer, 1998.

\bibitem{DBLP:conf/RTA/turuani'06}
Turuani M.
\newblock The cl-atse protocol analyser, 17th international conference on term
  rewriting and applications - rta 2006, seattle, wa/usa, july, 12, 2006.
\newblock 4098:277--286, 2006.

\bibitem{meadows96}
C.~Meadows.
\newblock The {NRL} protocol analyzer: an overview.
\newblock {\em Journal of Logic Programming}, 26(2):113--131, 1996.

\bibitem{Plandowski99}
W.~Plandowski.
\newblock Satisfiability of word equations with constants is in pspace.
\newblock In {\em FOCS}, pages 495--500, 1999.

\bibitem{Plandowski04}
W.~Plandowski.
\newblock Satisfiability of word equations with constants is in pspace.
\newblock {\em J. ACM}, pages 483--496, 2004.

\bibitem{Plandowski98}
W.~Plandowski and W.~Rytter.
\newblock Application of lempel-ziv encodings to the solution of words
  equations.
\newblock In {\em ICALP}, pages 731--742, 1998.

\bibitem{Sas}
Y.~Sasaki, Y.~Naito, N.~Kunihiro, and K.~Ohta.
\newblock Wang's sufficient conditions of md5 are not sufficient, 2005.
\newblock \url{http://eprint.iacr.org/}.

\bibitem{Schulz90}
K.~U. Schulz.
\newblock Makanin's algorithm for word equations - two improvements and a
  generalization.
\newblock In K.~U. Schulz, editor, {\em IWWERT}, volume 572 of {\em Lecture
  Notes in Computer Science}, pages 85--150. Springer, 1990.

\bibitem{WangFLY04}
X.~Wang, D.~Feng, X.~Lai, and H.~Yu.
\newblock Collisions for hash functions md4, md5 , haval-128 and ripemd.
\newblock \url{http://eprint.iacr.org/}, 2004.

\bibitem{WangLFCY05}
X.~Wang, X.~Lai, D.~Feng, H.~Chen, and X.~Yu.
\newblock Cryptanalysis of the hash functions md4 and ripemd.
\newblock In Cramer \cite{DBLP:conf/eurocrypt/2005}, pages 1--18.

\bibitem{WangYY05a}
X.~Wang, Y.~L. Yin, and H.~Yu.
\newblock Finding collisions in the full sha-1.
\newblock In V.~Shoup, editor, {\em CRYPTO}, volume 3621 of {\em Lecture Notes
  in Computer Science}, pages 17--36. Springer, 2005.

\bibitem{WangY05}
X.~Wang and H.~Yu.
\newblock How to break md5 and other hash functions.
\newblock In Cramer \cite{DBLP:conf/eurocrypt/2005}, pages 19--35.

\bibitem{Yaj}
J.~Yajima and T.~Shimoyama.
\newblock Wang's sufficient conditions of md5 are not sufficient, 2005.
\newblock \url{http://eprint.iacr.org/}.

\end{thebibliography}

\appendix



\end{document}